\documentclass[prb,twocolumn,superscriptaddress]{revtex4}
\usepackage{epsfig}
\usepackage{amsmath}
\usepackage{amsfonts}
\usepackage{amssymb,mathrsfs}

\begin{document}

\title{Linear magnetoresistance in compensated graphene bilayer}

\author{G.Yu.~Vasileva}
\affiliation{Institut f{\"u}r Festk{\"o}rperphysik, Leibniz Universit{\"a}t Hannover, Appelstra{\ss}e 2, D-30167 Hannover, Germany}
\affiliation{Ioffe Institute, Polytechnicheskaya 26 , 194021 St. Petersburg, Russia}
\affiliation{\mbox{Peter the Great Saint-Petersburg Polytechnic University, Politechnicheskaya 29,
195251 St. Petersburg, Russia}}

\author{D.~Smirnov}
\affiliation{Institut f{\"u}r Festk{\"o}rperphysik, Leibniz Universit{\"a}t Hannover, Appelstra{\ss}e 2, D-30167 Hannover, Germany}

\author{Yu.L.~Ivanov}
\affiliation{Ioffe Institute, Polytechnicheskaya 26 , 194021 St. Petersburg, Russia}

\author{Yu.B.~Vasilyev}
\affiliation{Ioffe Institute, Polytechnicheskaya 26 , 194021 St. Petersburg, Russia}

\author{P.S.~Alekseev}
\affiliation{Ioffe Institute, Polytechnicheskaya 26 , 194021 St. Petersburg, Russia}

\author{A.P.~Dmitriev}
\affiliation{Ioffe Institute, Polytechnicheskaya 26 , 194021 St. Petersburg, Russia}

\author{I.V.~Gornyi}
\affiliation{Ioffe Institute, Polytechnicheskaya 26 , 194021 St. Petersburg, Russia}
\affiliation{Institut f{\"u}r Nanotechnologie, Karlsruhe Institute of Technology, 76021 Karlsruhe,
Germany}
\affiliation{\mbox{Institut f{\"u}r Theorie der Kondensierten Materie, Karlsruhe Institute of Technology, 76128 Karlsruhe, Germany}}

\author{V.Yu.~Kachorovskii}
\affiliation{Ioffe Institute, Polytechnicheskaya 26 , 194021 St. Petersburg, Russia}

\author{M.~Titov}
\affiliation{Radboud University Nijmegen, Institute for Molecules and Materials, NL-6525 AJ
Nijmegen, The Netherlands}

\author{B.N.~Narozhny}
\affiliation{\mbox{Institut f{\"u}r Theorie der Kondensierten Materie, Karlsruhe Institute of Technology, 76128 Karlsruhe, Germany}}
\affiliation{National Research Nuclear University MEPhI (Moscow Engineering Physics Institute), 115409 Moscow, Russia}

\author{R.J.~Haug}
\affiliation{Institut f{\"u}r Festk{\"o}rperphysik, Leibniz Universit{\"a}t Hannover, Appelstra{\ss}e 2, D-30167 Hannover, Germany}

\date{\today}

\begin{abstract}
  We report a nonsaturating linear magnetoresistance in
  charge-compensated bilayer graphene in a temperature range from 1.5
  to 150 K. The observed linear magnetoresistance disappears away from
  charge neutrality ruling out the traditional explanation of the
  effect in terms of the classical random resistor network model. We
  show that experimental results qualitatively agree with a
  phenomenological two-fluid model taking into account electron-hole
  recombination and finite-size sample geometry.
\end{abstract}

\maketitle

Classical magnetoresistance is a perfect tool for experimental studies
of multicomponent electronic systems \cite{pippard} where conventional
theory of electronic transport \cite{abrikosov} predicts a quadratic
dependence of the resistance on the weak applied magnetic field
followed by a saturation in classically strong fields. While most
materials do exhibit the quadratic behavior,\cite{footnote} there is a fast growing
number of experiments reporting observations of linear
magnetoresistance (LMR) in a wide variety of novel materials including
multilayer graphenes \cite{weber2015,liao2012,fried2010}, topological
insulators
\cite{pavlo2015,mole2015,ocke2015,zhang2013,gusev2013,wang2012prl,wang2012,shek2012},
Dirac \cite{feng2015,novak2015,nara2015, ghimire15,liang15} and Weyl \cite{shek2015,shekhar1506}
semimetals, transition-metal dichalcogenides \cite{zhao2015} as well
as in narrow-gap semiconductors \cite{hu2008} and three-dimensional (3D)
silver chalcogenides \cite{hus2002,xu1997}.

Semiclassical linear magnetoresistance has been predicted for 3D
metallic slabs with complex Fermi surfaces and smooth boundaries
\cite{azb,lak}, for strongly inhomogeneous, granular materials
\cite{pal}, and for compensated two-component systems with
quasiparticle recombination \cite{us1}. Purely quantum effects (and
screening of charged impurities) lead to LMR in zero-gap band systems
with linear dispersion in the case where all carriers belong to the
first Landau level \cite{abrikos,abrikos1998,klier}. In weak fields,
quantum interference in two-dimensional electron systems yields an
interaction correction \cite{zna} to resistivity that is linear in the
Zeeman magnetic field.

The extreme quantum limit of Refs.~\onlinecite{abrikos,abrikos1998}
has been realized in graphene \cite{fried2010} and in Bi$_2$Se$_3$
nanosheets \cite{wang2012prl}. The quantum theory was also reported
\cite{pavlo2015} to be applicable to the novel topological material
LuPdBi. The classical theory of Ref.~\onlinecite{pal} was recently
used to interpret the behavior of hydrogen-intercalated epitaxial
bilayer graphene \cite{weber2015}. It was argued that large samples of
epitaxial bilayer graphene contain a ``built-in mosaic tiling'' due to
the dense dislocation networks \cite{butz2014}, making it an ideal
material to realize the random network model of Ref.~\onlinecite{pal}.
At the same time, neither theory can explain LMR in homogeneous
topological insulators \cite{brink2013} and neutral two-component
systems \cite{mole2015,gusev2013,hu2008}.

In this paper we report results of a systematic experimental analysis
of magnetotransport in exfoliated bilayer graphene. Precisely at
charge neutrality, we have observed nonsaturating LMR in a wide range
of magnetic fields in Hall bars of widths   ${0.5}$, ${0.95}$, and ${2.0}\,\mathrm{\mu m}$
in a temperature range from $1.5$ to $150\,\mathrm{K}$. Deviations from charge
neutrality lead to eventual saturation of the magnetoresistance. Our
key experimental findings are not accounted for within the random
resistor network model \cite{pal}. Indeed, this model is insensitive
to the relative concentration of different types of charge carriers
and thus cannot explain the observed saturation of the
magnetoresistance away from charge neutrality. This model also does
not explain the transition between the quadratic dependence at very
weak magnetic fields and LMR observed at higher fields
\cite{mole2015}. The extreme quantum limit is unlikely to be reached
in our system at ${150}\,\mathrm{K}$ for both electrons and holes
\cite{mole2015,brink2013}. Moreover, the excitation spectrum in
bilayer graphene is quadratic, which rules out the quantum theory of
Refs.~\onlinecite{abrikos,abrikos1998}.

We are able to explain our results in terms of a semiclassical
description of finite-size, charge-compensated two-component systems
in moderately strong, classical magnetic fields \cite{us1,mole2015}.
The key element of the physical picture of Ref.~\onlinecite{us1} is
the electron-hole recombination \cite{rb}. When external magnetic
field is applied, recombination processes allow for a neutral
quasiparticle flow in the lateral direction relative to the electric
current \cite{geim2013}. Although such neutral current cannot be
directly detected in our measurements, its presence leads to
redistribution of charge carriers over the sample area influencing the
nonuniform profile of the electric current in the sample. As a result,
the sample is essentially split into the bulk and edge regions, which
contribute to the total sheet resistance of the sample as parallel
resistors. The bulk and edge resistances exhibit qualitatively
different dependence on the magnetic field yielding LMR. Away from
charge neutrality a nonzero Hall voltage is formed leading to the
observed saturation of the magnetoresistance.

\section{Experimental details and sample characterization}
\label{exp}

\begin{figure}[t]
\centering
\includegraphics{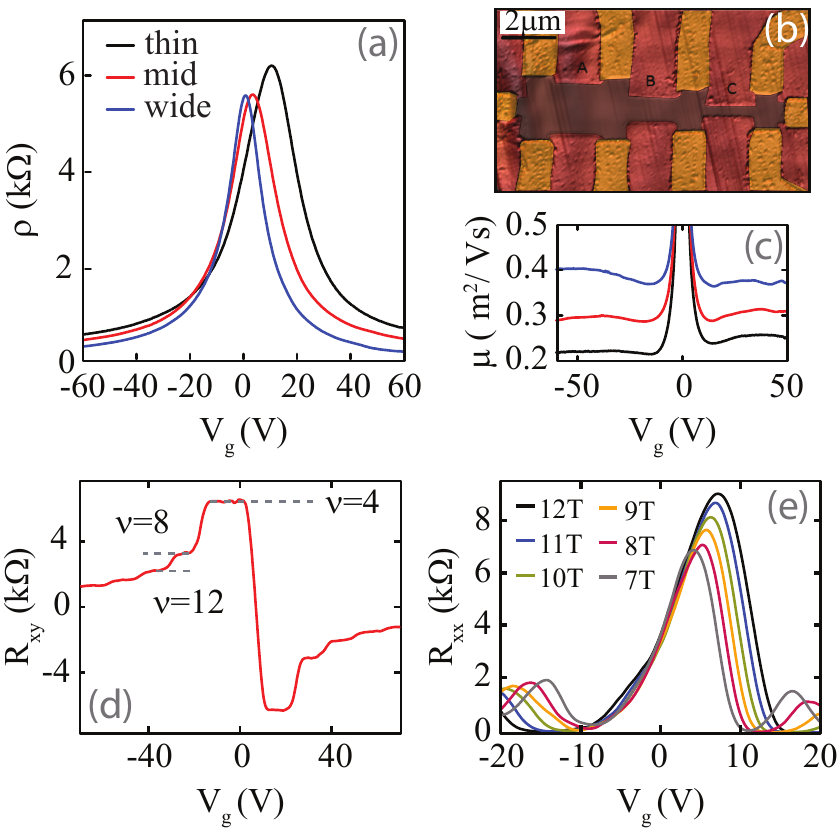}
\caption{(Color online) (a) The gate voltage dependence of the resistivity and (c) mobilities measured at B=$0\,\mathrm{T}$ and T=$25\,\mathrm{K}$  for the three sections of the device. (b) The AFM image of the sample (grey) with contacts (yellow). The sample contains three Hall bar sections $2$, $0.95$, and $0.5\,\mathrm{\mu m}$ wide (left to right). (d) Hall and (e) the longitudinal resistances in the wide section of the sample at T=$1.5\,\mathrm{K}$.}
\label{fig:sample}
\end{figure}

We have prepared the sample by placing the exfoliated bilayer graphene
sheet on the substrate consisting of a highly doped Si wafer covered
by a $330$nm-thick SiO$_2$ film. Subsequently, the sample was
patterned into a triple Hall bar device, see Fig.~\ref{fig:sample}(b)
for an atomic force microscope (AFM) image. The sample consists of
three sections $2$, $0.95$, and $0.5\,\mathrm{\mu m}$ wide. The length
of each Hall bar is $1.8\,\mathrm{\mu m}$. The sample was purified
using an AFM tip (instead of annealing) which allowed us to decrease the
concentration of the charged impurities on top of graphene
considerably. The carrier concentration $n$ in the sample can be
varied up to ${5\times10^{12}}$cm$^{-2}$ by applying a gate voltage
$V_{g}$ to the conducting substrate, which acts as a back gate.

Magnetotransport was studied by four-probe method with simultaneous
measurements of longitudinal $R_{xx}$ and transverse $R_{xy}$
resistances in perpendicular magnetic fields from $0$ to $7\,\mathrm{T}$ and in a
temperature range from $1.5$ to $150\,\mathrm{K}$ passing an {\it ac} current
with an amplitude of $10\,\mathrm{\mu A}$ through the sample.

To characterize the sample and to define the charge neutrality point
(CNP), the field effect (FE) was measured for each section of the
device. Figure~\ref{fig:sample}(a) shows the FE dependences measured at
${B=0\,\mathrm{T}}$ and ${T=25\,\mathrm{K}}$ for the three sections of the device.
All three sections exhibit a graphene typical FE with a sharp maximum corresponding to the CNP.
The precise value of $V_{g}^*$
corresponding to CNP depends on the Hall bar width and is shifted from
${0.8}\,\mathrm{V}$ in the widest section of the sample toward ${3.6}\,\mathrm{V}$ in the
medium and ${10.4}\,\mathrm{V}$ in the narrowest Hall bar. The maximum resistivity
in the widest and middle sections is $5.6\,\mathrm{k \Omega}$ while exceeding
$6.2\,\mathrm{k \Omega}$ for the narrowest section.

The electron and hole mobilities were estimated from the conductivity
at $25\,\mathrm{K}$ using the one-band model (see Fig.~\ref{fig:sample}(c)). The electron and hole densities necessary for this
estimate were obtained from the measured Shubnikov-de Haas
oscillations at low temperatures. The resulting mobilities increase
with the width of the sample; we have obtained the following values
for the mobilities of the narrowest, medium, and widest
sections of the sample far away from charge neutrality: $2200$,
$3000$, and $4000\,\mathrm{cm^2/Vs}$ for holes and $2600$, $3000$, and
$3800\,\mathrm{cm^2/Vs}$ for electrons.



\begin{figure}[t]
\centering
\includegraphics{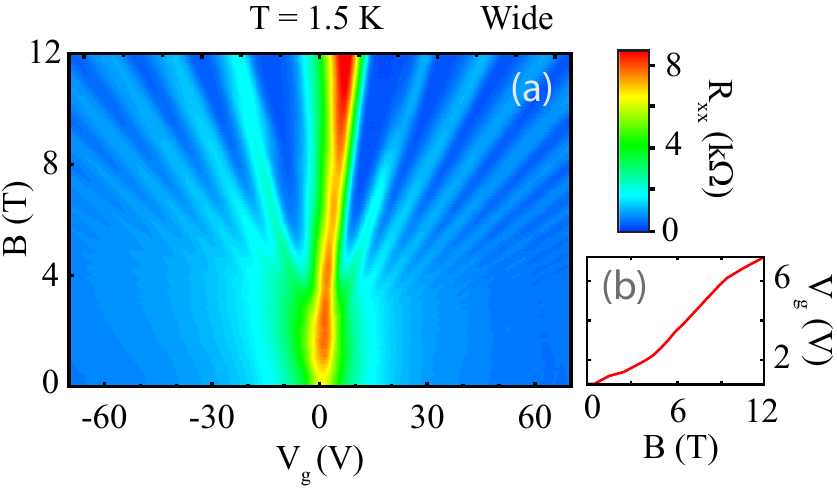}
\caption{(Color online) Color plot of $R_{xx}$ in the wide sample at
  {$T=1.5$}K as a function of magnetic field and gate voltage (a). The
  fan-like peak structure clearly demonstrates the Landau levels. The
  central peak shows the shift of the charge neutrality point with
  magnetic field (b).}
\label{fig:fan}
\end{figure}

The dependence of the mobility on the width of the sample is
attributed to scattering of carriers on the sample edges and is
described in Ref.\onlinecite{geomerty}. Although the above mobilities
are not very high, the samples are of a good quality having a clear
manifestation of CNP and exhibiting the quantum Hall effect (see
Fig.~\ref{fig:sample}(d)). Measurements of the Hall resistance in the
wide section of the sample at $1.5\,\mathrm{K}$ in relatively high
magnetic field ${9.5}\,\mathrm{T}$ (Fig.~\ref{fig:sample}(d))
demonstrate the features inherent to bilayer graphene following from
filling factors in the Hall plateaus equal to
${\nu=\pm4,\pm8,\pm12}$. In strong magnetic field the neutrality point
is shifted towards higher gate voltages, see Figs.~\ref{fig:sample}(e)
and \ref{fig:fan}. For the wide section of the sample at $12$T, CNP
corresponds to ${V_{g}=7}\,\mathrm{V}$. This effect has also been
observed in other sections of the sample.

\section{Linear magnetoresistance}

We have measured the longitudinal resistance for all three sections of
the sample and the Hall resistance between widest and medium sections
in the interval of gate voltages from $-20$ to $32.2\,\mathrm{V}$ with
the step ${\delta{V}_{g}=2.4}\,\mathrm{V}$ that includes CNP for all
three sections. The data for the wide section of the sample at
{$T=1.5$}K are shown in Fig.~\ref{fig:fan}. To reduce the conductance
fluctuations, further measurements were performed at higher
temperatures: $25$, $50$, $100$, and $150\,\mathrm{K}$. At such high
temperatures quantum effects, e.g. Landau quantization, are not
detectable.

\begin{figure}[t]
\centering
\includegraphics{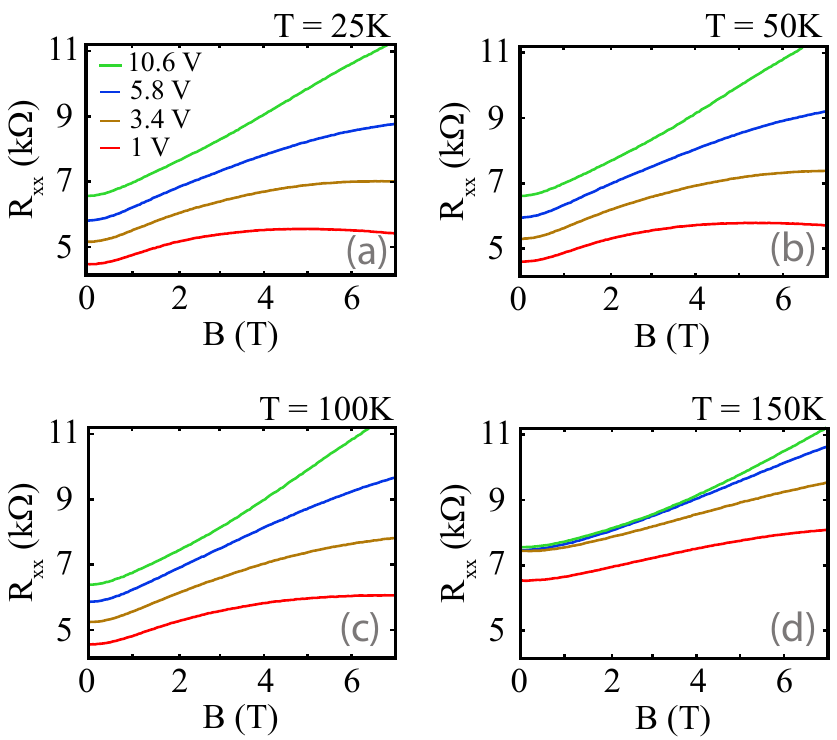}
\caption{(Color online) Magnetoresistance of the thin section of the
  sample at (a) $25$, (b) $50$, (c) $100$, and (d) $150\,\mathrm{K}$ for several gate voltages indicated on the plot.}
\label{fig:data1}
\end{figure}

\begin{figure}
\centering
\includegraphics{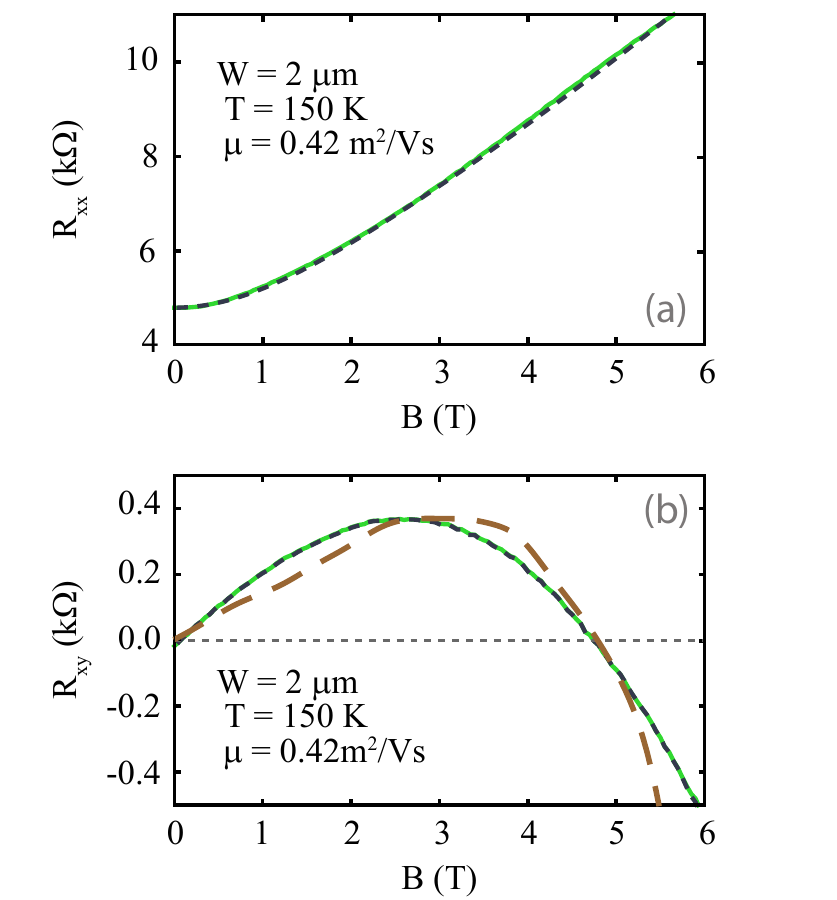}
\caption{(Color online) (a) Magnetoresistance of the wide section of
  the sample at ${T=150}\,\mathrm{K}$ and ${V_g=1}\,\mathrm{V}$,
  closest to the charge neutrality point. The solid (green) line
  represents the experimental data; the dashed (blue) line represents
  the theoretical fit using the semiclassical description adapted from
  Ref.~\onlinecite{us1}, see Eqs.~(\ref{theory}), with the parameters
  given in Table~\ref{table}. (b) Hall resistance of the wide section
  of the sample at ${T=150}\,\mathrm{K}$ and ${V_g=1}\,\mathrm{V}$.
  The solid (green) line represents the experimental data; the dashed
  (blue) line represents a fit by the theory (\ref{theory}) where the
  carrier density was obtained from the experimental values of
  $R_{xy}/R_{xx}(B)$; the brown curve shows the theoretical fit where
  the carrier density was recalculated from the observed dependence of
  the maximum resistance (i.e. CNP) on the magnetic field, see
  Figs.~\ref{fig:sample} and \ref{fig:fan}.}
\label{fig:wide}
\end{figure}

The magnetoresistance data for the thin section of the sample at the
four temperatures are shown in Fig.~\ref{fig:data1}. The data show
linear behavior close to the neutrality point (the green curve
corresponding to the gate voltage ${V_g=10.6}\,\mathrm{V}$). Away from
neutrality, the data show linear behavior for an intermediate range of
magnetic fields followed by a saturation at stronger fields. Similar
results were obtained for the other two sections of the sample. At the
same time, the Hall resistance grows in amplitude in strong fields,
see Fig.~\ref{fig:wide}.

Although these observations are in good qualitative agreement with the
theoretical predictions of Ref.~\onlinecite{us1}, there are several
additional factors that may (and probably do) conspire to yield the
observed behavior. The charge neutrality point in our samples shifts
towards higher gate voltages in high magnetic field. As a result, the
same value of $V_g$ may correspond to different carrier densities in
low and high fields. Shifting away from charge neutrality can cause
both the saturation of $R_{xx}(B)$, as exhibited by most curves in
Fig.~\ref{fig:data1}, and the nonzero Hall resistivity, see
Fig.~\ref{fig:wide}(b). The latter can also be due to electron-hole
asymmetry in the sample, where the mobilities of electrons and holes
are sufficiently different \cite{CNPshift,shift2}. Both effects may
appear if the sample contained macroscopic inhomogeneities or resonant
impurities, that strongly modify the density of states near charge
neutrality. Finally, in contrast to the theory worked out in
Ref.~\onlinecite{us1}, the length of our samples is comparable to
their width and the samples cannot be considered infinitely long.

\begin{table}[tbp]
\caption{Microscopic parameters obtained from analyzing the
  experimental data with the theory (\ref{theory}) for the three
  sections of the sample, see Figs.~\ref{fig:wide} and \ref{fig:mid}.}
\begin{ruledtabular}
\begin{tabular}{cccc}
& narrow & medium & wide \\
\hline\noalign{\smallskip}
W & ${0.52\mu}$m & ${0.95\mu}$m & ${2\mu}$m \\
$\mu$ & ${0.25}$m$^2$/Vs & ${0.35}$m$^2$/Vs & ${0.42}$m$^2$/Vs \\
$\ell_0$ & ${0.43\mu}$m & ${0.79\mu}$m & ${1.2\mu}$m \\
\end{tabular}
\end{ruledtabular}
\label{table}
\end{table}

\begin{figure} 
\centering
\includegraphics{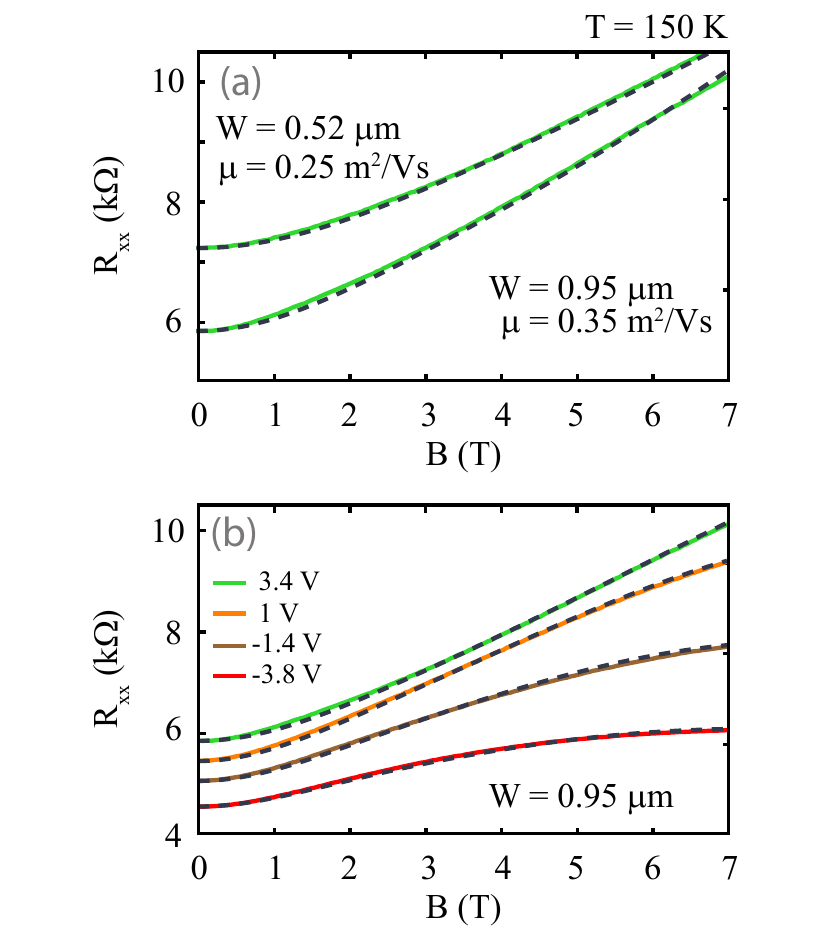}
\caption{(Color online) (a) Magnetoresistance of the narrow (top
  curves) and medium (bottom curves) sections of the sample at
  $150\,\mathrm{K}$ and the gate voltage closest to the charge
  neutrality point (${V_g=8.2}\,\mathrm{V}$ and
  ${V_g=3.4}\,\mathrm{V}$, respectively). The solid (green) lines
  represent the experimental data; the dashed (blue) lines represent
  the theoretical fit using the semiclassical description adapted from
  Ref.~\onlinecite{us1} to our sample geometry, see Table~\ref{table}
  for the complete set of parameters. (b) Magnetoresistance of the
  medium section of the sample at $150\,\mathrm{K}$ for several values
  of the gate voltage showing the onset of saturation as the system is
  tuned away from charge neutrality.}
\label{fig:mid}
\end{figure}

Some of the above complications present a significant challenge for an
analytic theory. Nevertheless, we may attempt to analyze the measured
data with the help of the existing theory of Ref.~\onlinecite{us1}. The
simplest version of this theory (applicable to a particle-hole
symmetric system with parabolic dispersion and energy-independent
impurity scattering rate) yields the following expressions for the
longitudinal and Hall resistivities of a two-component system near
charge neutrality:
\begin{subequations}
\label{theory}
\begin{equation}
\label{rxx}
R_{xx} = R_0 \frac{1+\mu^2 B^2}
{1+\mu^2 B^2\left[\frac{\tanh(W/\ell_R)}{W/\ell_R}\left(1-\frac{n^2}{\rho^2}\right)
+ \frac{n^2}{\rho^2}\right]},
\end{equation}
\begin{equation}
\label{rxy}
R_{xy} = \frac{R_0n}{\rho} \frac{(1+\mu^2 B^2)\mu B}
{1+\mu^2 B^2\left[\frac{\tanh(W/\ell_R)}{W/\ell_R}\left(1\!-\!\frac{n^2}{\rho^2}\right)
\!+\!\frac{n^2}{\rho^2}\right]}.
\end{equation}
Here $n$ and $\rho$ are the charge and quasiparticle densities, $\mu$
is the mobility (which is assumed to be the same for both electrons
and holes), $W$ is the sample width, $R_0$ is the zero-field
resistivity, and $\ell_R$ is the field-dependent recombination length.
Assuming that the dominant recombination process is the
impurity-assisted electron-phonon coupling that can occur anywhere in
the sample with equal probability, the recombination length found
in Ref.~\onlinecite{us1} is given by
\[
\ell_R = \frac{\ell_0}{\sqrt{1+\mu^2 B^2}}, \qquad \ell_0 = 2\sqrt{D\tau_R},
\]
where $D$ is the diffusion coefficient and $\tau_R$ is the
recombination time in zero magnetic field. As a result, in classically
strong fields (${\mu{B}\gg1}$) and for ${W\gg\ell_R}$ the
magnetoresistance (\ref{rxx}) close to the charge neutrality point is
linear, ${R_{xx}\approx{R}_0{W}\mu{B}/\ell_0}$.

Using the measured parameters of our sample in the above expressions,
we find that the theory predicts a magnetoresistance that is stronger
than what is actually observed in our experiment. However, our results
can be quantitatively described by Eqs.~(\ref{rxx}) and (\ref{rxy}) if
we introduce an empiric expression for the recombination length
\begin{equation}
\label{lr}
\ell_R = \frac{\ell_0}{\sqrt{1+\tilde\mu^2 B^2}},
\end{equation}
\end{subequations}
with ${\tilde\mu<\mu}$. This modification turns out \cite{kinur} to
effectively account for the following issues: (i) electron-hole
asymmetry, (ii) energy dependence of the electron-hole recombination
length and mobility, and (iii) spatial inhomogeneity of the sample. In
Figs.~\ref{fig:wide} and \ref{fig:mid} we used
${\tilde\mu\approx0.5\mu}$.

Electron-hole asymmetry manifests itself in the nonzero Hall
resistivity at charge neutrality. Moreover, for any value of the
carrier density the Hall resistivity is a nonmonotonous function of
the magnetic field. As mentioned above, this effect also leads to 
the apparent drift of the charge neutrality point (in terms of the
applied gate voltage) with the external magnetic field.

At temperatures lower than the Debye energy the dominant recombination
process involves electrons and holes near the bottom of the
band. Indeed, far away from the neutrality point, kinematic
constraints preclude the ``direct'' process where an electron from the
upper band is scattered into an empty state in the lower band by means
of single acoustic phonon emission. Instead, such ``hot'' electrons
require an additional scatterer (e.g. an impurity \cite{reiz} or a
second phonon) for recombination to take place. In contrast, electrons
close to the neutrality point in bilayer graphene are slow enough so
that the direct, single-phonon recombination is allowed. Hence, within
the kinetic equation approach \cite{kinur} the effective length scale
describing the recombination processes depends on energy. Similarly,
the impurity scattering time or carrier mobility is strictly speaking
energy-dependent as well. Now, the macroscopic description of
Ref.~\onlinecite{us1} involves quantities that are averaged over the
quasiparticle spectrum. Taking into account the existence of the
several distinct recombination processes, we arrive at the conclusion
that after thermal averaging, the typical recombination length
$\ell_R$ may be described by slightly different effective parameters
as compared to, e.g., Drude conductivity.

The width dependence of the carrier mobility indicates that the edge
region of the sample is characterized by stronger scattering. As a
result, all parameters describing electronic transport acquire an
effective coordinate dependence across the sample. Since in strong
magnetic fields the current is mostly flowing near the sample edges
\cite{us1}, we expect that the effective recombination length $\ell_R$
is determined by the lower mobility typical of the near-edge region.

Theoretical results shown in Figs.~\ref{fig:wide} and \ref{fig:mid}
were obtained by using expressions (\ref{theory}) with the parameters
listed in Table~\ref{table}. The theory (\ref{theory}) assumes that
electrons and holes have the same mobility. While plotting
Figs.~\ref{fig:wide} and \ref{fig:mid} we have treated the mobility as
a free parameter instead of using the values quoted in Sec.~\ref{exp}
(see Fig.~\ref{fig:sample}(a)) since experimentally one can reliably
determine mobility only far away from the neutrality point, where the
classical Hall resistivity exhibits the standard behavior
${R_{xy}={B}/(ne)}$. Close to charge neutrality, the mobility
may deviate from such experimental values due to electron-hole
interaction processes similar to the drag effect. Far away from charge
neutrality this interaction is ineffective since with exponential
accuracy only one band is partially filled and contributes to
low-energy physics. In contrast, close to the neutrality point, both
electrons and holes participate in transport and hence one has to take
into account their mutual scattering. The resulting change of the
mobility does not exceed 50\% in accordance to theoretical
expectations.

The values of the recombination length $\ell_0$ shown in
Table~\ref{table} show significant dependence on the sample width
(roughly, ${\ell_0\sim{W}}$). We interpret this observation as an
indication of a much larger recombination length that would
characterize a very large (in theory -- infinite) sample (if it were
possible to fabricate without strong structural disorder
\cite{weber2015,butz2014}). Assuming that the electron-hole
recombination is dominated by electron-phonon interaction (either
impurity- or edge-assisted), we argue that in narrow samples the
phonon spectrum is modified (compared to an idealized infinite
system), leading to a much shorter recombination length of the order
of the sample width.

The Hall resistance (\ref{rxy}) is expected to vanish at the
neutrality point. However, as we have already mentioned, in our sample
the neutrality point shifts toward higher gate voltages when a strong
magnetic field is applied. In order to account for this effect, we
have used the ratio ${R_{xy}/R_{xx}=\mu{B}n/\rho}$ to extract the
field-dependent quantity ${\mu{B}n/\rho}$ from the experimental data.
Using thus obtained dependence in Eq.~(\ref{rxy}), we find good quantitative
agreement between the calculated and measured values, see
Fig.~\ref{fig:wide}. At the same time, the longitudinal resistivity
(\ref{rxx}) is much less sensitive to small deviations of
density. Using the extracted values of ${\mu{B}n/\rho}$ in
Eq.~(\ref{rxx}) does not lead to visible changes in the calculated
curve shown in Fig.~\ref{fig:wide}.

The shift of CNP with magnetic field was observed directly, see
Figs.~\ref{fig:sample} and \ref{fig:fan}. Assuming that the maximum of
the longitudinal resistivity corresponds to CNP, we can extract the
field dependence of the chemical potential (and hence, carrier
densities) from the data. Using thus obtained dependence, we
recalculated the Hall resistance, see the brown curve in
Fig.~\ref{fig:wide}(b). The result shows reasonable agreement with the
data, with the visible deviations may stem from the mismatch of
temperatures in the two data sets in Figs.~\ref{fig:fan} and
\ref{fig:wide} (${T=1.5}$K and ${T=150}$K, respectively).

Finally, away from the neutrality point the data shows a tendency
towards saturation in high magnetic fields, see Fig.~\ref{fig:mid}(b).
The theoretical fits where performed with a set of parameters
depending on the gate voltage and taking into account the shift of CNP
with magnetic field. In particular, the mobility appeared to show a
slight increase from ${0.35}$m$^2$/Vs (close to CNP, see
Table~\ref{table}) to ${0.44}$m$^2$/Vs at ${V_g=-3.8}$V. At the same
time, in that range of gate voltages the recombination length $\ell_0$
appears to be almost unchanged from the value shown in
Table~\ref{table}).

\section{Conclusions}


In this paper we reported the experimental observation of linear
magnetoresistance in narrow bilayer graphene samples. The observed
behavior is in good qualitative agreement with the two-fluid model of
Ref.~\onlinecite{us1}. The observed effect is specific to the charge
neutrality point. Away from neutrality the magnetoresistance shows an
approximate linear behavior only in a limited intermediate range of
magnetic fields followed by a tendency to saturation. Our observations
are incompatible with the quantum theory of
Refs.~\onlinecite{abrikos,abrikos1998} and with the random resistor
network model of Ref.~\onlinecite{pal}, but are accounted for in the
semiclassical theory of two-component compensated systems of
Ref.~\onlinecite{us1}.

%

Using an empirical modification of the simplest theoretical model
(\ref{theory}), we were able to describe our data in a quantitative
fashion. A microscopic theory accounting for the physics that is
beyond the simplest version of the two-fluid model of
Ref.~\onlinecite{us1} should be based on the quantum kinetic equation
\cite{kinur,us2}. Further aspects of the phenomenon of linear
magnetoresistance are the subject of future experimental work,
especially in numerous novel materials.

\begin{acknowledgments}
\noindent
We are grateful to U. Briskot, M. Dyakonov, A.D. Mirlin,
M. Sch{\"utt}, and S. Wiedmann for helpful discussions. This work was
supported by the Dutch Science Foundation NWO/FOM 13PR3118, DFG SPP
1459, GIF, the EU Network Grant FP7-PEOPLE-2013-IRSES ``InterNoM'',
the Humboldt Foundation, the Russian Foundation of Basic Research (grant No. 15-02-04496-a, 14-02-00198 A), the Dynasty Foundation, the Grant of Russian
Ministry of Education and Science (Contract No. 14.Z50.31.0021), and
the President Grant for Leading Scientific Schools NSh-1085.2014.2.
\end{acknowledgments}


\begin{thebibliography}{39}
\bibitem{footnote}
	In wide samples of graphene monolayers a square-root
	magnetoresistance has been observed in Ref.~\onlinecite{mono-exp} near the 
	Dirac point, resulting from the combination of linear dispersion of carriers and short-range character of scatterers~\cite{mono-theo}.
\end{thebibliography}
\bibliographystyle{unsrt}

\end{document}